\documentclass[letterpaper, 10 pt, conference]{ieeeconf}  

\IEEEoverridecommandlockouts                              

\usepackage{graphics}
\usepackage{hyperref}
\usepackage{stfloats}
\usepackage{ifthen}
\usepackage{amsmath}
\usepackage{amssymb}
\usepackage[normalem]{ulem} 
\usepackage[colorinlistoftodos, textwidth=2.5cm]{todonotes}
\usepackage{multirow}
\usepackage{booktabs}
\usepackage{microtype}
\usepackage{url}
\usepackage[T1]{fontenc}
\usepackage{cite}

\usepackage{tikz}

\newcommand{\pie}[1]{%
\begin{tikzpicture}
 \draw (0,0) circle (1ex);\fill (1ex,0) arc (0:#1:1ex) -- (0,0) -- cycle;
\end{tikzpicture}%
}

\newboolean{showcomments}
\setboolean{showcomments}{true} 
\ifthenelse{\boolean{showcomments}}
  {
		\newcommand{\nbb}[2]{
		\fcolorbox{black}{yellow}{\bfseries\sffamily\scriptsize#1}
		{\sf$\blacktriangleright$\textcolor{blue}{\textit{#2}}$\blacktriangleleft$}
		}
		
		\newcommand{\remarks}[1]{\color{red}[#1]\color{black}}

		\newcommand{\del}[1]{\textcolor{red}{\sout{#1}}} 
  }
  {
		\newcommand{\nbb}[2]{}
		\newcommand{\remarks}[1]{}

		\newcommand{\del}[1]{} 
  }

\title{\LARGE \bf
Paving the Roadway for Safety of Automated Vehicles:\\ An Empirical Study on Testing Challenges}

\author{Alessia Knauss$^{1}$, Jan Schr{\"o}der$^{2}$, Christian Berger$^{2}$, and Henrik Eriksson$^{3}$
\thanks{*This work was supported by Vinnova grant 2014-06229.}
\thanks{$^{1}$Alessia Knauss is with the Department of Computer Science and Engineering,
Chalmers University of Technology, Gothenburg, Sweden
  {\tt\small alessia.knauss@chalmers.se}}%
\thanks{$^{2}$Jan Schr{\"o}der and Christian Berger are with  the Department of Computer Science and Engineering, University of Gothenburg, Sweden
        {\tt\small \{christian.berger, jan.schroder\}@gu.se}}%
\thanks{$^{3}$
Henrik Eriksson is with the Department of Electronics, RISE Research Institutes of Sweden, Bor\r{a}s, Sweden 
        {\tt\small henrik.eriksson@sp.se}}%
}

\begin{document}

\maketitle
\thispagestyle{empty}
\pagestyle{empty}

\begin{abstract}
The technology in the area of automated vehicles is gaining speed and promises many advantages. However, with the recent introduction of conditionally automated driving, we have also seen accidents. 
Test protocols for both, conditionally automated (e.g., on highways) and automated
vehicles do not exist yet and leave researchers and practitioners with different challenges. For instance, current test procedures do not suffice for fully automated vehicles, which are supposed to be completely in charge for the driving task
and have no driver as a back up.
This paper presents current challenges of testing the functionality and safety of automated vehicles derived from conducting focus groups and interviews with
26 participants from five
countries having a background related to testing automotive safety-related topics. 
We provide an overview of the state-of-practice of testing active safety features as well as challenges that needs to be addressed in the future to ensure safety for automated vehicles. 
The major challenges identified through the interviews and focus groups, enriched by literature on this topic are related to 1)~virtual testing and simulation, 2)~safety, reliability, and quality, 3)~sensors and sensor models, 4) required scenario complexity and amount of test cases, and 5) handover of responsibility between the driver and the vehicle.
\end{abstract}

\section{Introduction}

Today's technological advancements in the automotive domain bring more and more automation to our vehicles. The size of vehicle software is growing exponentially, while the safety has also to increase to allow for driving without a driver as a back up. 
Cases like Toyota's unintended acceleration~\cite{Koopman14} show that already trivial mistakes
can cause severe fatalities. Furthermore, the recent introduction of conditionally automated
driving has also resulted in several accidents~\cite{GoogleCrash, TeslaCrash, NutonomyCrash}.

These examples show that testing needs to cover a broad spectrum of unforeseeable situations and characteristics for automated vehicles where the driver is not monitoring the driving activity any more and cannot serve as a fall back option. 
Standardized test protocols, such as EuroNCAP, exist to systematically test
features aiming to support the driver with safety functionality to avoid or mitigate
accidents.  
However, such standards do
not exist for automated driving. 
In this paper, we investigate the following research question:\\ \textit{\textbf{RQ:} What are the challenges that have to be
addressed in order to test (conditionally) automated vehicles?}

Based on focus groups with practitioners from Sweden as well as interviews with practitioners and researchers from Sweden, Germany, the US, Netherlands, and China, we systematically gather and discuss testing challenges that need to be
overcome in the near future to ensure safety of increasingly automated vehicles. 
The 26 participants are from eight different companies (e.g., five of them premium
automotive OEMs), seven research institutes and universities, and one proving ground. We have presented an excerpt of the identified challenges for the software engineering community in \cite{Knauss2017}. 
In this paper, we analyze and identify the gap between current state-of-the-practice and future needs for testing of increasingly automated vehicles, highlight and explain all identified areas where research and development are urgent and most beneficial. Additionally, the identified areas are validated against open literature which of today is quite limited.

The paper is structured as follows: In Section \ref{sec:background} we provide background
information and discuss related work. Section \ref{sec:research-methodology} outlines our
research methodology. We present our results on the state-of-the-practice of safety-related
aspects in Section \ref{sec:results-state-of-practice} and on challenges in testing
automated vehicles in Section \ref{sec:results-future}. 
In Section \ref{sec:threats}, we depict threats to validity and conclude our paper in Section \ref{sec:conclusion}.


\section{Background and Related Work}
\label{sec:background}
\subsection{Background} A number of active safety systems has been developed to avoid or mitigate the
consequences of common accident scenarios such as: rear-end, on-coming, and run-off
road. To find representative scenarios and define viable test
methods, accident data has been studied. To perform the resulting rating
tests, an advanced driving robot is needed in the test vehicle
 to achieve repeatability and the necessary
precision and accuracy. Additionally, a soft target 
having its own propulsion system will act as sensor stimulus
for the function under test~\cite{seiniger2016test}.

In specific use cases, such as traffic jams or highway driving, lateral and longitudinal
functionality are combined to reach a higher level of automation by simultaneous
control of steering and acceleration/deceleration. To be able to discuss the different challenges at
different levels of automation ranging from no to full automation, different standardization
bodies have defined a limited number of levels and their meaning.  In this paper we use the
definition of automation levels proposed by the Society of Automotive Engineers (SAE) \cite{saej3016}. 
We define automated driving as SAE levels 3-5.   

\subsection{Related Work:} The success of automated road vehicles depends on several aspects; testing, verification, validation, and certification being some of them~\cite{schoner2016challenges,fahrenkrog2016technical,agaram2016validation,mazzega2016testing,wachenfeld2016release,paulweber2017validation}. Existing work indicates that relevant test scenarios and systematic evaluation approaches need to be defined for the assessment of automated vehicles. Furthermore, since field tests become unviable
due to the large number of kilometers to be 
driven~\cite{wachenfeld2016release,kalra2016driving}, the majority of the testing
must be performed using simulations or virtual test driving~\cite{pfeffer2016continuous},
whose results are validated on proving grounds or in field tests. Virtual testing can
also contribute to identify a further relevant test scenarios~\cite{sippl2016simulation}, and the
number of tests to perform can be reduced using combinatorial testing~\cite{wotawa2017testing}.
Test tracks need to complement testing with orchestrating several automated actors involved in a scenario~\cite{schoner2016testing}. In highly automated vehicles, the
hand-over~\cite{morgan2016handover} between driver and vehicle and vice versa
becomes crucial~\cite{schoner2016challenges} in order to avoid issues such as mode
confusion and unfair transitions~\cite{johansson2016safetransitions}.


\section{Research Methodology} 
\label{sec:research-methodology}
In our exploratory empirical investigation we used two different instruments of data collection: First degree and third degree data sources. First degree included \textit{focus groups and interviews} \cite{SSS08} (Section \ref{sec:focusgroup}), while third degree included studying \textit{existing research publications} (Section \ref{sec:literature-analysis}) related to the studied topic.

\subsection{Focus Groups and Interviews}
\label{sec:focusgroup}
First, we used focus groups to enable a rich discussion and a broad understanding of the studied topic, before focusing on specific details. The interviews aimed at understanding the topics in more detail.

\subsubsection{Participants}
Due to the sensitive topic studied in this paper, many participants were concerned taking part in this study and share their insights. Hence, to maximize the response rate, we used convenience sampling \cite{SSS08} and invited only our contacts in the area of active safety testing from industry and academia. Participants from academia were chosen to have a close collaboration with industry, and being involved in industry projects related to automated vehicles.

\textit{\textbf{Focus groups:}} 
First, we conducted \textit{four focus groups with eleven participants} from Sweden between September $7^{th}$, 2015 to November $25^{th}$, 2015. Focus group 1 included five engineers and one manager with up to 15 years experience in safety testing. 
Focus  group 2 consisted of one manager and one senior researcher with 19 and 12 years experience in automotive safety. Focus group 3 included two engineers and one manager with up to 14 years experience, and focus group 4 one manager with over 30 years experience in automotive safety. The last focus group could also be classified as interview -- as the rest of the participants canceled their participation.
The focus groups included representatives from one supplier, one proving ground for active safety testing and related researcher, and two automotive OEMs.

\textit{\textbf{Interviews:}}
We conducted \textit{15 (semi-)structured interviews with participants from five different countries}. Interviews were conducted to include each participant's view individually in our study and took place between May $23^{rd}$, 2016 and October $24^{th}$, 2016. 
Table~\ref{tab:participants-interviews} depicts the details of our interview participants: country, position, and years of experience in an area related to automotive safety.
Participants are split into research (i.e., four participants currently employed in research institutes and three at universities) and industry (i.e., seven participants from automotive OEMs and one from an automotive supplier), sorted based on years of experience.

The response rate for the interviews was 60\%. We have approached 25 possible participants, eleven from Germany, six from Sweden, six from US, one from China, and one from Netherlands. We were able to conduct 15 interviews. Furthermore, due to data concerns of the participants, we present the results anonymously, do not share the transcripts, and do not describe the employment details of the participants, but only present their position and years of expertise.

\begin{table}[t]
\caption{Overview of interview participants (research and industry)}
\label{tab:participants-interviews}
\centering
\small
\begin{tabular}{ l  c c c }
\hline
\textbf{ID} & \textbf{Country} & \textbf{Position} & \textbf{Experience} 
\\\hline\hline
\multicolumn{2}{l}{\textbf{Research}}&&
\\\hline
R.1 & Germany & Manager & 20 years 
\\\hline
R.2 & Netherlands & Manager & 16 years 
\\\hline
R.3 & Sweden & Manager & 15 years 
\\\hline
R.4 & Sweden & Manager & 10 years 
\\\hline
R.5 & Germany & Manager & 9 years 
\\\hline
R.6 & China & Researcher & 6.5 years 
\\\hline
R.7 & Sweden & Manager & 6 years 
\\\hline\hline
\multicolumn{2}{l}{\textbf{Industry}}&&
\\\hline
I.1 & Sweden & Manager & 30 years 
\\\hline
I.2 & Sweden & Engineer \& Researcher & 12 years 
\\\hline
I.3 & Germany & Manager & 10 years 
\\\hline
I.4 & Germany & Engineer & 7 years 
\\\hline
I.5 & US & Engineer & 6 years 
\\
\hline
I.6 & Germany & Manager &5 years 
\\\hline
I.7 & Germany & Manager & 5 years 
\\\hline
I.8 & Germany & Engineer \& Researcher & 4 years 
\\\hline
\end{tabular}
\end{table}

\subsubsection{Data Collection}
Each focus group and interview contained two parts, part 1) focusing on the state-of-the-practice of testing active safety systems, and part 2) on future trends. The interview lengths was 45 minutes, the focus groups 60 minutes in lengths to allow for enriched discussions.  Four interviews were conducted in person, the rest took place on Skype or phone.

Due to the exploratory nature of the study, we only used a few open-ended questions to not restrict the participants going into one specific direction. The major question used for part 1) state-of-the-practice was ``How do you test active safety systems?'', while for part 2) we used ``What are the future trends on vehicle automation (looking at SAE Levels 3-5)? How will testing have to change to address these future trends topics?''.
Part 1) and 2) contained the same sub-questions:
\begin{itemize}
\item What are the stakeholders of testing active safety systems/automated vehicles?
\item What are the quality criteria for successfully testing active safety systems/automated vehicles?
\item What are the processes, methods, and tools involved in testing of active safety systems/automated vehicles?
\end{itemize}

Each focus group and interview was conducted by two researchers. The first author had always the role of a moderator,
asking questions and giving directions to the participants to explore the topic under study. The second researcher was responsible for taking notes and was alternating either the second or third author or Hang Yin, another researcher from a similar research field. 

\subsubsection{Data Analysis}
To analyze the collected data and derive our findings, we applied techniques recommended for empirical research methods \cite{SSS08}: We transferred the notes from both the focus groups and the interviews into a list of over 1000 separate statements consisting of 1-3 sentences that logically belong together. Hence, the content of the focus groups was handled with the same weight as one interview due to the fact that we could not track back the contribution of each participant in the focus groups.  The individual results and further details on the focus groups (e.g., results from applying word frequency analysis on the notes) can be found in Knauss et al.~\cite{Knauss2016}. 

For the list of statements, we applied coding starting from statements provided by the four focus groups, and continuing with the statements from the interviews. First we assigned low-level codes and later allocated them to respective high-level clusters. We assigned topics to these clusters and iterated once again to make sure that the codes are assigned properly. We present the topics with example statements in Section \ref{sec:results-state-of-practice} for state-of-the-practice and Section \ref{sec:future-trends-focusgroups} for future trends. 

\subsection{Analysis of Challenges in Related Research Publications}
\label{sec:literature-analysis}
At the beginning of our study in August 2015 there were hardly any research publications about systematic testing for autonomously driving vehicles targeting commercial applications. This initiated our exploratory research method using participants views. While we were conducting our study several related papers have been published. Hence, the analysis of related literature is considered as a final step, comparing our results to their results. We identified ten papers that are considered to be the most relevant ones for testing of automated vehicles~\cite{schoner2016challenges}-\cite{wotawa2017testing}. We compare whether and which of the challenges we identified from our interviews and focus groups are presented in these ten publications. We present these results in Section \ref{sec:literature}.

\section{Results: State-of-the-practice of Testing Active Safety Systems}
\label{sec:results-state-of-practice}
The following results are based on the focus groups and interviews. They are meant to provide the necessary information on the stakeholders, current quality criteria, processes, methods, and tools to understand the main contribution of the paper regarding challenges of testing automated vehicles.

The \textit{\textbf{stakeholders}} of testing active safety systems mentioned by the participants of this study sorted by priority are: OEMs and their different subgroups (e.g., developers, feature owners, project managers, testers), suppliers, Government and organizations for test methods/catalogs definition/legislative organizations, customers, certification bodies running different tests, proving grounds, others like researchers, insurance companies and journalists.

The identified \textit{\textbf{processes and methods}} are:
\begin{itemize}
\item \textit{Development processes} following for example the V-model, reaching from requirements elicitation, to test definition, and implementation of active safety functions. Testing during development reaches from testing of lower level separate components to full vehicle testing. This includes software, but also hardware-related testing: electronic control unit (ECU) testing, testing of functionality, model-in-the-loop, software-in-the-loop, hardware-in-the-loop, and vehicle-in-the-loop testing, etc.~as well as integration tests on the test track. Agile methodologies are also identified by the participants as a commonly used development practice, for which contract-based design is used where contracts are defined between the components and can be tested against. 
\item \textit{Proving ground testing} plays an important role in active safety testing (e.g., used for release testing). However, OEMs try to minimize the testing on the proving ground as it is costly. For release testing, certain scenarios are tested for, which are currently derived from analyzing accident data and finding the most common accident patterns. 
\item Instead, \textit{simulation and virtual reality testing} are used before going to the proving ground to make sure the functions work under different conditions. Again, simulation and virtual reality are used as a means to test the identified scenarios from accident data. Data from real test drives are used to simulate driving scenarios.
\item \textit{Certification testing} based on norms like EuroNCAP.
\item Finally, the vehicles are \textit{running on public roads} -- currently these test drives are a means to collect real data to be used for testing in simulations rather than used as a testing technique. 
\end{itemize}

The identified \textit{\textbf{quality criteria}} (i.e., criteria that are used to determine whether a test is successful) span different levels of development related testing including: concrete quality attributes, use case testing, testing of sensor related aspects, and testing whether certain scenarios are fulfilled. 
If an OEM follows the V-model, there are certain tests that are part of the formal process and are derived from the requirements for verification and validation.
For testing of the final vehicle, collision avoidance and criteria in predefined (certification) tests represent quality criteria. An active safety system is supposed to actively avoid accidents or mitigate the consequences thereof. Hence,  collision avoidance or the remaining velocity during a crash are one of the mentioned quality criteria. For certification testing, quality criteria are exactly defined (e.g., in EuroNCAP, ISO 26262 certification). 

The \textit{\textbf{tools}} in-use identified by the participants are:
\begin{itemize}
\item Related to \textit{proving ground testing}: Driving robots, test targets (e.g., soft targets) used to simulate vehicles or other targets (e.g., pedestrians) as well as the target carriers, cameras or eye trackers (e.g., to measure the driver behavior), reference positioning system, rapid prototyping tools, data logging and collection tools, and automatic reporting tools.
\item \textit{Development tools}: Requirements management tools, risk estimation tools, HMI development tools, HIL, scripting tools, IDE, code generation, databases, validation and verification tools, sensor interface replacement tools.
\item \textit{Modeling and simulators}: Simplified models of vehicles and traffic sets, as well as modeling of accident scenarios are needed. Furthermore, sensor models (including noise of sensors like rain and fog) have to be created to simulate the sensor input during simulation testing.
\item \textit{Data related tools}: Sensor fusion tools, data collection tools (e.g., CAN bus data in vehicles) and labeling of data tools, databases and analysis of logged data.
\item A few \textit{tools to automate testing} tasks were mentioned like test creation and analysis of results. 
\end{itemize}


\section{Results: Challenges of Testing Automated Vehicles}
\label{sec:results-future}
\subsection{Identified Challenges in Focus Groups and Interviews}
\label{sec:future-trends-focusgroups}
\begin{figure}[t]
\centering
\includegraphics[width=0.46\textwidth]{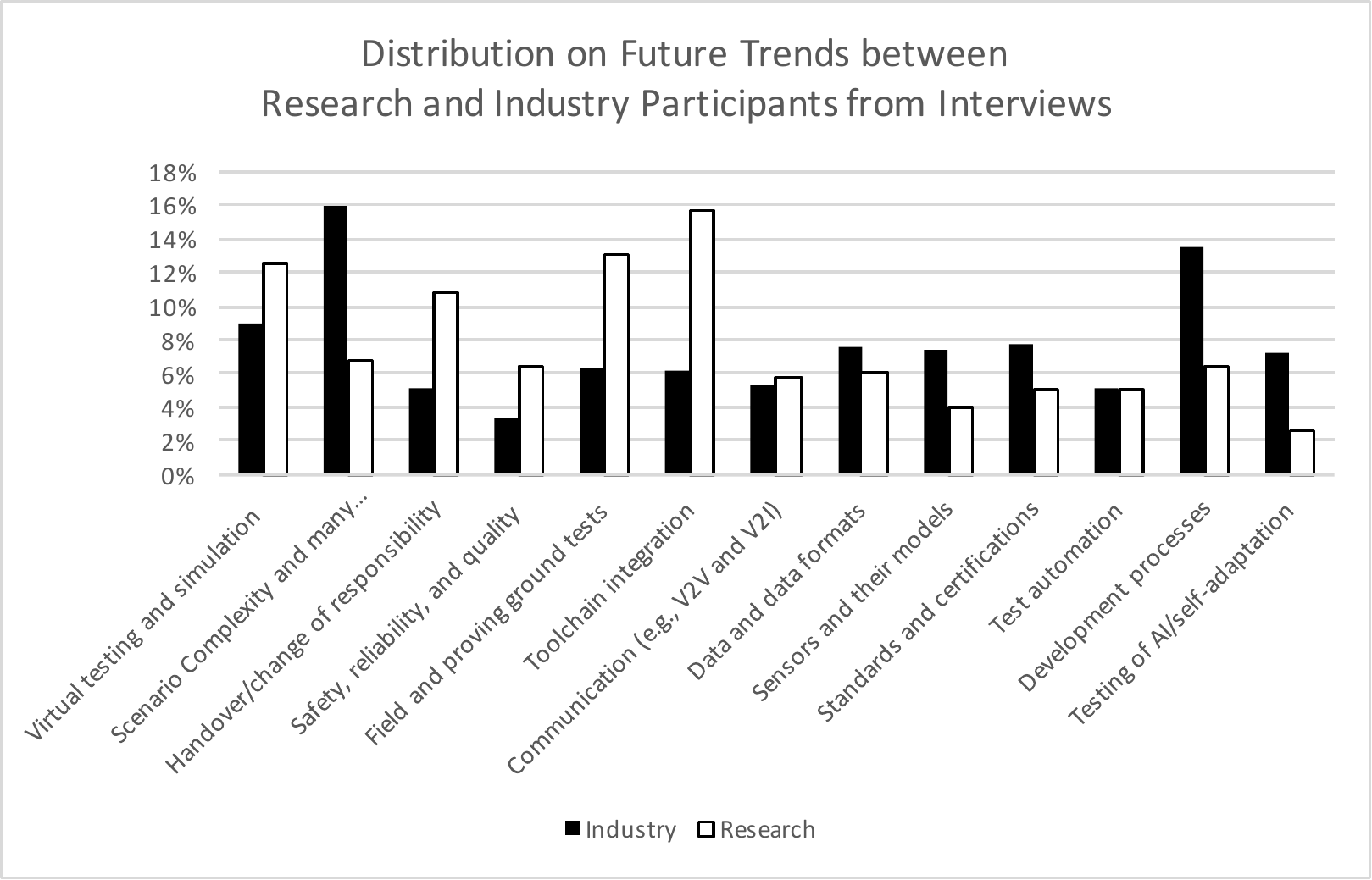}
\caption{Overview of identified challenges by industry and research based on the interviews. The y-axis represents the distribution of the statements from industry/research related to this challenge in comparison to all notes from industry/research. Industry distribution is visualized in black, research distribution in white.}
\label{fig:summary-future-research-industry}
\end{figure}

A summary of challenges when testing automated vehicles is presented in Fig.~\ref{fig:summary-future} based on the content of the four focus groups and the 15 interviews. The challenges are sorted based on the amount of participants discussed the corresponding challenge. Figure \ref{fig:summary-future-research-industry} depicts the challenges with their distribution of participants from research and industry from the interviews. We describe each of the challenges in detail:

\begin{figure}[h!]
\centering
\includegraphics[width=0.46\textwidth]{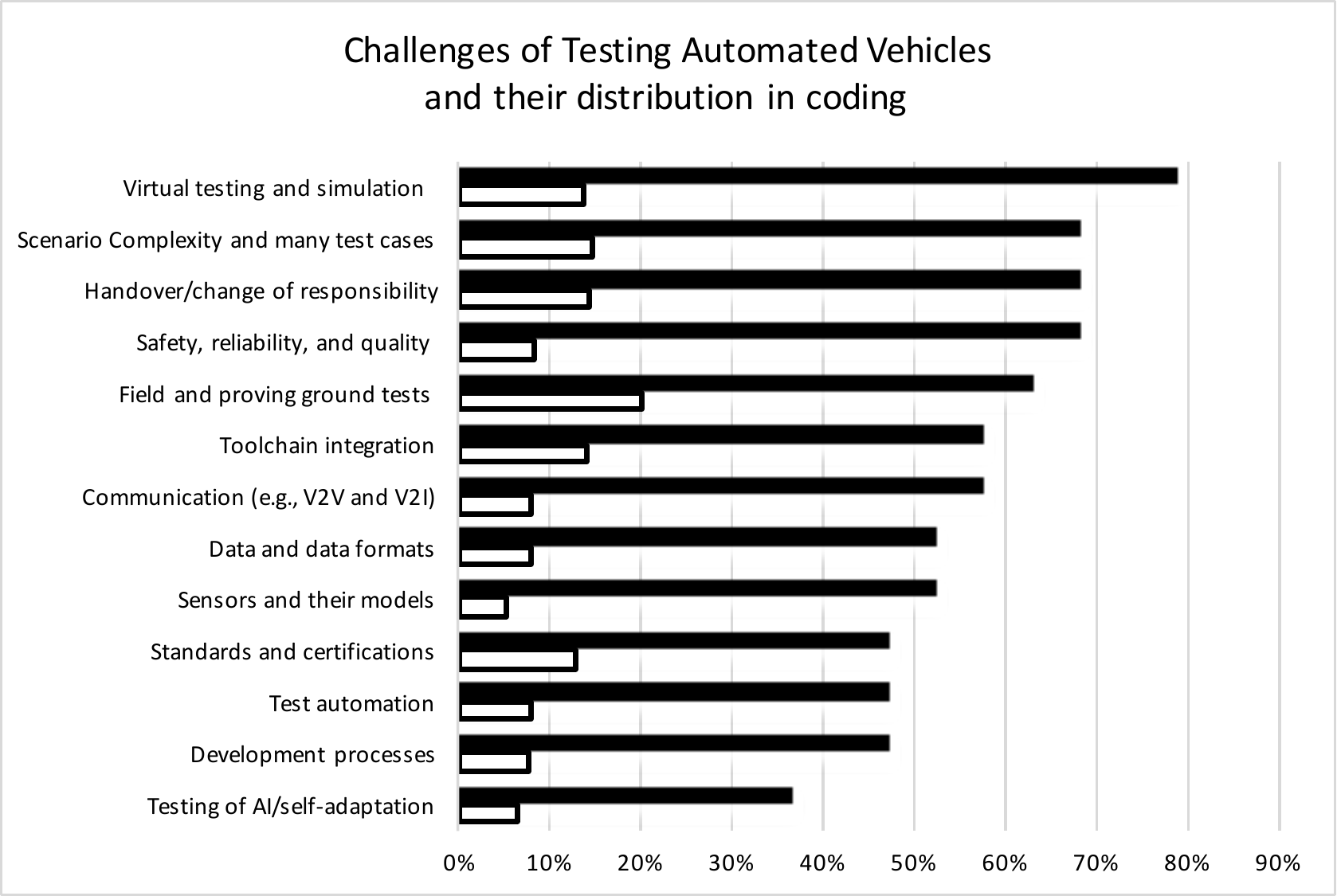}
\caption{Overview of all identified challenges and their distribution \cite{Knauss2017} -- percentage of participants that mentioned the topic in black and percentage of how many of the notes contained that topic in white.}
\label{fig:summary-future}
\end{figure}

\textbf{\textit{Virtual testing and simulation -- increasingly important:}}
Because of the vast amount of testing that is required for automated vehicles and the high costs for practical testing, virtual testing and simulation gain increased importance as indicated by the participants. 
The advantages of virtual testing are 
\begin{itemize}
\item  efficient testing, as it can support parallel testing or during day and night, which a proving ground cannot support. Hence, testing the vehicle for the required millions of kilometers ``can be reached over a couple of days''. 
\item allows to recreate complex traffic scenarios using real traffic data. It is difficult to recreate scenarios in exactly the same way on a proving ground without simplifying the scenarios. Hence, simulation testing allows to test with more data and different conditions, considering different weather, climate, and driving conditions in different areas.
\item supports testing of human-related aspects. Simulation and virtual reality allow for a safe testing environment, compared to proving ground or real life testing, especially considering that ``people [might] use the system differently than expected?''
\item possible to combine virtual reality and simulation with field testing. This augmented reality can be used to test complex traffic scenarios, using the real data communication between the vehicles and the infrastructure. 
\end{itemize}  

Despite the advantages of simulation and virtual reality testing, there are many challenges to be addressed:
\begin{itemize}
\item Can certifications be based on simulations or virtual reality? The legal aspects of using simulations as a means to achieve certifications need to be explored by the testing authorities. 
\item If testing in simulation and real environments is combined (i.e., mixed reality testing), a specific infrastructure is required. 
\item Better simulation tools regarding test efficiency.
\item Virtual vehicles and accurate sensor models.
\item Benchmark scenarios for simulations.
\item A simulation environment relies on real-world data collected from real traffic. For each scenario the right data set needs to be identified and field data needs to be collected.  
\item Automatic labeling of data is needed to allow the application of artificial intelligence.
\end{itemize}

\textbf{\textit{Scenario complexity and many test cases -- necessary to test for:}}
 At the moment, complex scenarios with multiple test objects cannot be tested at the proving ground, but will become important for automated vehicles. ``Complexity unexpectedly high'': As automated vehicles will go everywhere and have to be tested that they are safe in every environment, not only the complexity of the scenarios but also the variety of test cases to test for will have to increase. ``Almost impossible to think about all possible traffic scenarios.'' If not all test cases are possible to test for at the proving ground, virtual reality and simulation are suggested as one possible solution. Then the tests can run in parallel and can run non-stop. More data, kilometers, and testing different conditions as well as considering different weather,  climate, and driving conditions in different areas is necessary to consider during testing to guarantee safety in all possible conditions. 
Furthermore, machine learning will be necessary to cope with the data. 

\textbf{\textit{Change of responsibility -- handover increases non-functional aspects testing:}}
An active safety system is only supporting the driver. For automated vehicles there is a shift from driver in charge to vehicle in charge where the vehicle has to be fully available at all times, making right decisions and making sure the passengers are safe. Many challenges arise due to this shift.  

For conditionally automated vehicles, testing includes scenarios where the vehicle hands over the driving task to the driver, but also the driver might want the responsibility back: ``how to test an automated vehicle with test people sitting inside the car and the vehicle might `go bananas'? ''.
In cases where the vehicle cannot handle the driving task (e.g., due to failures in the vehicle), it needs to be able to give back the responsibility to the driver: ``Testing that the human can take over in case the vehicle requests to''. One solution might be that a vehicle should always be able to go in safe mode, in case the driver cannot take over. Testing for this criteria on ``Does a car go to safe mode?'' is crucial.  
 
With the change of responsibility due to increased vehicle automation there is a need to consider a broader range of testing non-functional properties.   
While for the lower levels 3-4 the testing of driver awareness is important, for full automation car-sickness is one of the interviewee's concerns. Similarly, in lower levels of automation the vehicle should postpone its action to as late as possible to give the driver the chance to react. In fully automated vehicles 
the vehicle reaction (e.g.\ braking) is enabled as early as possible to support driver comfort (e.g.\ non-functional testing). 
In an automated vehicle you have to trust the system and feel safe as your safety depends on the automated vehicle. Based on the interviews, testing for the safety feeling poses a challenge. User-friendliness and robustness of functions are other non-functional requirements to consider testing for. 

\textbf{\textit{Safety, reliability, and quality -- increasingly important:}}
There are different aspects on how automated vehicles will make driving safer, e.g. through a centralized speed control on highways, preventing cars from speeding. Traffic jams will decrease, due to central control. Despite the advantages that a central instance will bring, each of the vehicles itself must be safe. People that are supposed to use automated vehicles expect vehicles to not be involved in accidents: ``There will be zero-failure acceptance: customers want to have bullet-proof cars''. However, other opinions are that ``It does not have to be perfect. It should be better than humans, but how good does it have to be to be better than humans?'' Regulations should define how safe safe enough is: ``The question is: is 100 life safe enough if 2 are killed?'' Higher safety standards and stricter validation \& verification are needed. Certification will play a major role. Legal aspects focus on defining when a certain quality is reached: ``How do you make sure that I do not miss an object in front of me?'' 

Quality has to increase, while it is still impossible to test the vehicles in all possible conditions. Reliability and accountability are also mentioned as major criteria for automated vehicles. Concerning testing of automated vehicles, the testing on proving grounds should be safe in respect to test equipment and processes, especially when testing fully automated vehicles which might behave in unexpected ways.

\textbf{\textit{Field and proving ground tests -- have their limitations:}}
For Level 4 and 5, there will be even more sensors and communication of vehicles. This requires new equipment on proving grounds, the targets should be remote controlled individually from one central instance to ensure that certain predefined situations are guaranteed, and new test methods in respect to automation of test processes should be developed: ``Everything will be similar but much harder''. Data logging from many entities, test scenarios that are less predictable, interactions with other vehicles, altering weather and environmental conditions are challenging topics mentioned for proving ground testing. 

As EuroNCAP will not be sufficient for automated vehicles, more complex scenarios will be used for Level 4-5 vehicles. It will be less deterministic, needs to cover more kilometers, consider more data, different conditions, weather, and climate. Limitations of proving grounds are: limited roads, limited space, light conditions, not possible to test all scenarios. Certain aspects might be tested though continuous experimentation. Recorded real world traffic data can be collected with this method to also work offline with this data. However, continuous experimentation  has its limitations (e.g.\ privacy aspects). Hence, the proposed solution is to test certain aspects, especially early phases, on the test track, and otherwise move to virtual testing or real world testing. 

\textbf{\textit{Tool chains -- integration of different testing techniques necessary:}}
New, well-defined, and established test processes are required for automated vehicles. The manual testing established for non-automated vehicles will require an increase of software engineers if the same procedures will be applied to automated vehicles. To counteract this costly aspect, (semi-)automation of test processes as well as the use of cheaper testing techniques (e.g., simulation testing) than real-life testing is needed. One example that was mentioned was support to automatically identifying scenarios for major real-world events. 

For the increasing levels of automation, scenarios are becoming more complex. For full automation the costs of testing all functionality on the proving ground are too high. It seems crucial to test certain functionality beforehand. Furthermore, in automated vehicles not only the triggering of the functionality needs to be tested. Additionally, it has to be investigated whether a certain functionality is not triggered when it should not, monitored over a recommended distance~\cite{kalra2016driving}. 
 
Fully automated vehicles will change their behavior at run-time. Real-world testing will be inevitable in these cases. As not all requirements are known when developing an automated system and features will be added after the vehicle is delivered, real-world testing will allow continuous software engineering. Proving ground testing will then be used at development time. 

\textbf{\textit{Communication (e.g., vehicle-to-infrastructure, vehicle-to-vehicle) -- support necessary to ensure availability:}} Communication between everything, and the infrastructure to support this communication is crucial for future automated vehicles. The infrastructure has to provide interfaces for automated vehicles: ``it's about understanding each others intentions/states'' and testing whether correct messages are sent. IT companies supporting this communication will be key players in this ecosystem. Availability -- a major characteristic of automated vehicles, will be supported through redundant communication, and could be enabled by vehicle-to-vehicle and vehicle-to-infrastructure communication.

\textbf{\textit{Data and data formats -- are becoming increasingly important:}}
For automated vehicles it is necessary to use data from real-driving in different conditions -- not only very specific predefined scenarios as is the state-of-the-practice for active safety systems. The collection, analysis, labeling, and providing data in a systematic and efficient way is a crucial point for testing of automated vehicles. Predefined data formats are necessary to reuse data for different tools. Furthermore, the amount of data that is collected for every drive considering the amount of drives we need for later reuse and analysis represents a big data problem. Hence, data providers that focus on providing data in a systematic way and serve with input data together with databases, but also provide some kind of automation of the test evaluations were mentioned as a new type of stakeholders.

Another important aspect concerns the data from real test drives and on how to transfer the huge amount of data in an efficient and secure way (e.g., V2V, test drives). Testing on whether the data that a vehicle received from another entity (e.g., parking provider) is correct and the vehicle could trust this entity is necessary to implement in automated vehicles. 

\textbf{\textit{Sensors and their models:}}
In non-automated driving the driver is the instance that monitors everything and reacts to certain situations. In automated vehicles, the vehicle itself must ``see everything'' and is not supposed to ever fail. Hence, topics like sensor availability, performance, validation, and redundancy mechanisms are important. Questions arise like: 
How do the sensors perform? What is the field of view for the sensors? Are there redundant mechanisms implemented? 

To answer these questions, lots of data from real traffic is needed. Further sensors and better sensors with dedicated tool chains are needed to be able to simulate all maneuvers. Sensor settings have to be recalibrated efficiently, sensor data has to be compared to ground truth data, concerning accurate positioning, time, and synchronization of vehicle data.

During simulation and virtual reality testing certain sensors need to be simulated. Sensor models are used for this purpose. These models are supposed to correctly replace real sensors to run on real computers in real time and should represent the real environment. Hence, these models should also model faulty behavior: ``How detailed a world model has to be is difficult to say.'' 

\textbf{\textit{Standards and certifications:}}
To guarantee robustness of automated vehicles, the definition and introduction of quality criteria, regulations, standards, and certifications are required. 
Participants mentioned that new insurance companies and policy makers might be introduced to deal with this. An approach similar to the star system (e.g., EuroNCAP) for the current safety systems is also needed for automated vehicles for each of the Levels 3-5. Other options to explore for certification include virtual testing or testing in real traffic. 
Furthermore, it is unclear how the certification of system functionality that is realized through machine learning algorithms will be achieved, which might result in potential re-certification. 

\textbf{\textit{Test automation -- required to increase efficiency of testing:}}
Artificial Intelligence applied on big data is one promising strategy to deal with the complexity of automated vehicles and especially with their testing. With increased automation levels the amount of testing required to guarantee safety has to be expanded. Test automation will support efficient testing. Examples for test automation mentioned by the study participants include: 
\begin{itemize}
\item Automated data analysis applied for different problems where patterns need to be found in big data. For example, new processes are needed to automatically extract test cases from real-world events. 
\item Automated data collection and attribute labeling. 
\item Automated testing on proving grounds. 
\item Automated testing integrated into fully automated vehicles to guarantee safety at runtime. 
\end{itemize}

\begin{table*}[h!]
\caption{Ranking of challenges based on our study (from Figure \ref{fig:summary-future-research-industry}) as well as based on the ranking of statements of the same challenges in literature. An average ranking is calculated and results are sorted based on this average rating.}
\label{tab:challenges}
\centering
\begin{tabular}{ l  l  c  c || c }
\hline
\textbf{Challenge} & \textbf{Literature} & \textbf{Ranking} & \textbf{Ranking} & \textbf{Ranking (Lit. } \\ 
& & \textbf{(Literature)} & \textbf{(Our study)}& \textbf{\& our study)}\\ \hline\hline
Virtual testing and simulation & \cite{schoner2016challenges,fahrenkrog2016technical,agaram2016validation,mazzega2016testing,wachenfeld2016release,paulweber2017validation,pfeffer2016continuous,sippl2016simulation,wotawa2017testing} & \pie{324} & \pie{284} & \pie{304}\\ \hline
Safety, reliability, and quality & \cite{schoner2016challenges,agaram2016validation,mazzega2016testing,wachenfeld2016release,paulweber2017validation,kalra2016driving,wotawa2017testing}& \pie{252}& \pie{244}& \pie{248}\\ \hline
Sensors and their models     & \cite{schoner2016challenges,mazzega2016testing,agaram2016validation,wachenfeld2016release,paulweber2017validation,pfeffer2016continuous}& \pie{216}& \pie{190}& \pie{203}\\ \hline
Scenario complexity     & \cite{agaram2016validation,wachenfeld2016release,pfeffer2016continuous,sippl2016simulation} & \pie{144}& \pie{244}  & \pie{194}\\ \hline
Handover/change of responsibility~ & \cite{schoner2016challenges,fahrenkrog2016technical,agaram2016validation,mazzega2016testing} & \pie{144}& \pie{244} & \pie{194}\\ \hline
Field and proving ground tests   & \cite{fahrenkrog2016technical,agaram2016validation,mazzega2016testing} & \pie{108}& \pie{227} & \pie{168}\\ \hline
Toolchain integration            & \cite{wachenfeld2016release,paulweber2017validation,pfeffer2016continuous} & \pie{108}& \pie{209}& \pie{159}\\ \hline
Test automation                 & \cite{fahrenkrog2016technical,agaram2016validation,pfeffer2016continuous,wotawa2017testing} & \pie{144}& \pie{171} & \pie{158}\\ \hline
Standards and certification     & \cite{schoner2016challenges,agaram2016validation,paulweber2017validation} & \pie{108}& \pie{171}& \pie{140}\\ \hline
Communication (e.g., V2I)                & \cite{paulweber2017validation}& \pie{36}& \pie{209} & \pie{123}\\ \hline
Data and data formats            & \cite{paulweber2017validation} & \pie{36}& \pie{190} & \pie{113}\\ \hline
Development processes            & \cite{pfeffer2016continuous} & \pie{36}& \pie{171} & \pie{104}\\ \hline
Test of AI/self-adaptation     & \cite{wotawa2017testing} & \pie{36}& \pie{133} & \pie{85}\\ \hline
\end{tabular}
\end{table*}

\textbf{\textit{Development processes -- flexibility necessary:}} Not all requirements are defined from the beginning in automated vehicles. Requirements need to be added while the vehicle is developed as well as updated after the vehicle is delivered. The development processes need to be adjusted to allow for a more flexible approach. During the development of the vehicle
the ability to move more efficiently between the different development activities (e.g., in the V-model) is needed. 
Experts assess that the life cycle will become shorter. Accordingly, the processes should allow for these shorter iterations. Thus, it is not clear how suitable the V-model will remain for this kind of development. Current processes might partially remain V-like, while other parts might need different approaches.
However, OEMs will only focus on sub-processes to handle the complexity of a fully automated vehicle. To support the full process of developing an automated vehicle, an increased number of software engineers is needed, which might be difficult to employ by a single OEM. Furthermore, test-driven approaches will gain importance, requiring to start testing at the beginning of the life cycle. This supports efficient processes, in which one get maximum amount of results with least amount of effort. Agile methodologies seem to provide the advantage of having testing in focus and allow for an increased quality with permanent testing activities. 

Due to the importance of real data in developing certain functionality, continuous software engineering, especially continuous experimentation will gain increased importance. During continuous experimentation, disabled functions are delivered with the vehicle that collect run-time data. However, there are some concerns with this method that have to be solved in order for this method to be acceptable: 1) the privacy concerns, and 2) because the actual actuators are disabled, it is not certain that a required feature would have triggered and hence cannot be used as a reliable testing technique. ``Testing is not supposed to ever stop'' to make sure that the changes in the system behavior are safe. As one of the major changes with fully automated vehicles is that the driver is not involved in the driving task and will not be required to monitor the driving of the vehicle.

\textbf{\textit{Testing of AI/Adaptation -- needs to be considered:}} As automated vehicles will be used in unobserved environments, the development of features needs to consider and will rely on data collected at runtime and will even use artificial intelligence to support system functionality.
With these techniques, automated vehicles will continuously learn and gain new knowledge. At certain points they will adapt their behavior, also known as self-adaptive capability. Hence, automated vehicles will not always behave deterministic, as we expect them to. During testing, automated vehicles might e.g.~learn what is right and what is wrong and pass a test without problems next time. This behavior, however, might only be triggered in a similar environment -- in this case the test environment, and not achieve the same results in a similar but different environment. 
Questions remain on how to test and validate (self-)adaptive behavior. Testing simple scenarios will not be sufficient. Complex scenarios which are less deterministic having more degrees of freedom are needed. 
Furthermore, the testing cannot stop after the vehicle is delivered to the customers. The vehicle has to be tested for learning effects. 
The current state-of-the-practice does not cover testing of machine learning and artificial intelligent techniques; novel processes, techniques, and tools are required to support this kind of testing.


\subsection{Priorities of Challenges enriched by Literature}
\label{sec:literature}
The results of analysing whether and how many recent publications contain challenges we identified is illustrated in Table~\ref{tab:challenges}.
The conclusion from the comparison is that all challenges identified in this paper are covered in at least one of the other papers. Virtual testing and simulation is the most common challenge in the literature and in our study. 
Similarly, development processes and testing of artificial intelligence/self-adaptation are the least common challenges. There are however a few discrepancies: sensor and sensor modeling as well as test automation are more common in literature than in our study. Contrary, V2X communication is a more common challenge in our study than in literature. 
Table \ref{tab:challenges} summarizes these results through an adjusted ranking, based on the average of both results: our study results and challenges covered in the 10 studied publications. 

\section{Threats to Validity}
\label{sec:threats}

The presented challenges are not necessarily related to new research areas, neither are they meant as an exhaustive representation of all existing challenges. Our work rather aims at matching the industrial and academic view to highlight important gaps. For example, to our surprise, the topic of security issues was not discussed but represents usually an important challenge. For some challenges presented even commercial approaches exist already but seem to have weaknesses. Furthermore, we acknowledge that we only focused on a few stakeholder groups. With this, we purposefully did not cover all stakeholders of this ecosystem.  For example, we could have included policy makers but focused specifically on the three stakeholder groups (i.e., OEMs, suppliers, and researchers), as the topic is already very broad and we would risk to loose our focus by including even further stakeholder groups.

Our data collection is based on four focus groups and 15 interviews. This amount might be considered too limited to draw conclusions from. However, due to the sensitivity of the data around automated vehicles, participants, especially from the premium manufacturers, are cautious to participate in this kind of studies. Our results mostly include participants in management positions and hence expected to have a good overview of their discipline. Furthermore, our results indicate a saturation for the identified challenges. We reached an agreement between at least 3 participants on each of the 13 identified challenges after conducting the focus groups and 6 interviews. After analyzing all collected data, we have at least an agreement of 7 participants for each challenge. Hence, we are confident that including further participants in the study would not have changed the results to a large extent; if at all, then just shifting the weights of the identified challenges. Another mitigation strategy that we applied was to additionally analyze related work for these challenges, which again confirmed our results, as all challenges identified by us could also be found in the studies published in the last year.

The data was analyzed by the first author and could introduce some biases to the results. We have taken a sample of the first five interviews, resembling 20\% of the entire statements from the focus groups and interviews. This sample was analyzed in terms of assigning codes to the content by the second author. Codes matching all 13 challenges have been identified, some on a lower level of abstraction. Additionally, the second author identified 2 additional topics. 

Another bias could be related to researchers taking notes instead of using transcripts. Due to the sensitivity of our research topic, we tried to mitigate the risk of participants not attending due to audio recording and transcribing the interviews afterwards. Hence, we decided to take notes during the interviews. Our mitigation strategy to not introduce biases from the person taking notes was to have three different researchers involved in taking the notes. Hence, this should even out preassumptions or directions for a favorable topic. 


\section{Conclusion and Future Work}
\label{sec:conclusion}
We have presented an empirical study on the challenges of testing automated vehicles. We have compared our results to the challenges identified in literature in 2016, as this topic starts to attract research attention. The major challenges identified are related to 1) virtual testing and simulation, 2) safety, reliability, and quality, 3) sensors and their models, 4) required scenarios complexity and amount of test cases, and 5) handover between driver and vehicle and shift of the responsibility to the vehicle.

Future work should extend this study to include participants from further countries, stakeholder groups, and focus on addressing the presented challenges.



\section*{Acknowledgments}
We are
deeply indebted to all participants of our study and Dr.\ Hang Yin for notes taking during several interviews. 

\addtolength{\textheight}{-12cm}   






\bibliographystyle{IEEEtran}
\bibliography{IEEEabrv,library}

\begin{thebibliography}{10}
\providecommand{\url}[1]{#1}
\csname url@rmstyle\endcsname
\providecommand{\newblock}{\relax}
\providecommand{\bibinfo}[2]{#2}
\providecommand\BIBentrySTDinterwordspacing{\spaceskip=0pt\relax}
\providecommand\BIBentryALTinterwordstretchfactor{4}
\providecommand\BIBentryALTinterwordspacing{\spaceskip=\fontdimen2\font plus
\BIBentryALTinterwordstretchfactor\fontdimen3\font minus
  \fontdimen4\font\relax}
\providecommand\BIBforeignlanguage[2]{{%
\expandafter\ifx\csname l@#1\endcsname\relax
\typeout{** WARNING: IEEEtran.bst: No hyphenation pattern has been}%
\typeout{** loaded for the language `#1'. Using the pattern for}%
\typeout{** the default language instead.}%
\else
\language=\csname l@#1\endcsname
\fi
#2}}

\bibitem{Koopman14}
\BIBentryALTinterwordspacing
P.~Koopman, ``A case study of toyota unintended acceleration and software
  safety,'' September 2014. [Online]. Available:
  \url{https://users.ece.cmu.edu/~koopman/pubs/koopman14_toyota_ua_slides.pdf}
\BIBentrySTDinterwordspacing

\bibitem{GoogleCrash}
\BIBentryALTinterwordspacing
{The Verge}, ``{Google Self-Driving Car Crash Report},'' 2016, last access
  September 15th, 2016. [Online]. Available:
  \url{http://www.theverge.com/2016/2/29/11134344/google-self-driving-car-crash-report}
\BIBentrySTDinterwordspacing

\bibitem{TeslaCrash}
\BIBentryALTinterwordspacing
{The Guardian}, ``{Tesla Self-driving Car -- Death during driving in
  Autopilot},'' 2016, last access September 15th, 2016. [Online]. Available:
  \url{https://www.theguardian.com/technology/2016/jun/30/tesla-autopilot-death-self-driving-car-elon-musk}
\BIBentrySTDinterwordspacing

\bibitem{NutonomyCrash}
\BIBentryALTinterwordspacing
{Channel NewsAsia}, ``{nuTonomy halts vehicle trials after accident at
  one-north},'' 2016, last access October 25th, 2016. [Online]. Available:
  \url{http://www.channelnewsasia.com/news/singapore/nutonomy-halts-vehicle-trials-after-accident-at-one-north/3219314.html}
\BIBentrySTDinterwordspacing

\bibitem{Knauss2017}
A.~Knauss, J.~Schroeder, C.~Berger, and H.~Eriksson, ``Software-related
  challenges of testing automated vehicles,'' in \emph{Proceedings of the
  International Conference on Software Engineering (ICSE)}, 2017.

\bibitem{seiniger2016test}
P.~Seiniger and A.~Weitzel, ``Test methods for consumer protection and
  legislation for {ADAS},'' \emph{Handbook of Driver Assistance Systems: Basic
  Information, Components and Systems for Active Safety and Comfort}, pp.
  213--230, 2016.

\bibitem{saej3016}
{SAE, J3016}, ``Taxonomy and definitions for terms related to on-road motor
  vehicle automated driving systems,'' 2014.

\bibitem{schoner2016challenges}
H.-P. Sch{\"o}ner, ``Challenges and approaches for testing of highly automated
  vehicles,'' in \emph{Energy Consumption and Autonomous Driving}.\hskip 1em
  plus 0.5em minus 0.4em\relax Springer, 2016, pp. 101--109.

\bibitem{fahrenkrog2016technical}
F.~Fahrenkrog \emph{et~al.}, ``Technical evaluation and impact assessment of
  automated driving,'' in \emph{Road Vehicle Automation 3}.\hskip 1em plus
  0.5em minus 0.4em\relax Springer, 2016, pp. 237--246.

\bibitem{agaram2016validation}
V.~Agaram \emph{et~al.}, ``Validation and verification of automated road
  vehicles,'' in \emph{Road Vehicle Automation 3}.\hskip 1em plus 0.5em minus
  0.4em\relax Springer, 2016, pp. 201--210.

\bibitem{mazzega2016testing}
J.~Mazzega \emph{et~al.}, ``Testing of highly automated driving functions,''
  \emph{ATZ worldwide}, vol. 118, no.~10, pp. 44--48, 2016.

\bibitem{wachenfeld2016release}
W.~Wachenfeld and H.~Winner, ``The release of autonomous vehicles,'' in
  \emph{Autonomous Driving}.\hskip 1em plus 0.5em minus 0.4em\relax Springer,
  2016, pp. 425--449.

\bibitem{paulweber2017validation}
M.~Paulweber, ``Validation of highly automated safe and secure systems,'' in
  \emph{Automated Driving}.\hskip 1em plus 0.5em minus 0.4em\relax Springer,
  2017, pp. 437--450.

\bibitem{kalra2016driving}
N.~Kalra and S.~M. Paddock, ``Driving to safety: How many miles of driving
  would it take to demonstrate autonomous vehicle reliability?''
  \emph{Transportation Research Part A: Policy and Practice}, vol.~94, pp.
  182--193, 2016.

\bibitem{pfeffer2016continuous}
R.~Pfeffer and T.~Leichsenring, ``Continuous development of highly automated
  driving functions with vehicle-in-the-loop using the example of {Euro NCAP}
  scenarios,'' in \emph{Simulation and Testing for Vehicle Technology}.\hskip
  1em plus 0.5em minus 0.4em\relax Springer, 2016, pp. 33--42.

\bibitem{sippl2016simulation}
C.~Sippl \emph{et~al.}, ``From simulation data to test cases for fully
  automated driving and {ADAS},'' in \emph{IFIP International Conference on
  Testing Software and Systems}.\hskip 1em plus 0.5em minus 0.4em\relax
  Springer, 2016, pp. 191--206.

\bibitem{wotawa2017testing}
F.~Wotawa, ``Testing autonomous and highly configurable systems: Challenges and
  feasible solutions,'' in \emph{Automated Driving}.\hskip 1em plus 0.5em minus
  0.4em\relax Springer, 2017, pp. 519--532.

\bibitem{schoner2016testing}
H.-P. Sch{\"o}ner and W.~Hurich, ``Testing with coordinated automated
  vehicles,'' \emph{Handbook of Driver Assistance Systems: Basic Information,
  Components and Systems for Active Safety and Comfort}, pp. 261--276, 2016.

\bibitem{morgan2016handover}
P.~Morgan \emph{et~al.}, ``Handover issues in autonomous driving: A literature
  review,'' 2016.

\bibitem{johansson2016safetransitions}
R.~Johansson \emph{et~al.}, ``Safe transitions of responsibility in highly
  automated driving,'' in \emph{Proceedings of The Ninth International
  Conference on Dependability, DEPEND 2016}.\hskip 1em plus 0.5em minus
  0.4em\relax IARIA, 2016, pp. 21--27.

\bibitem{SSS08}
\BIBentryALTinterwordspacing
F.~Shull \emph{et~al.}, Eds., \emph{{Guide to Adv. Empirical Software
  Engineering}}.\hskip 1em plus 0.5em minus 0.4em\relax London: Springer
  London, 2008. [Online]. Available:
  \url{http://www.springerlink.com/index/10.1007/978-1-84800-044-5}
\BIBentrySTDinterwordspacing

\bibitem{Knauss2016}
A.~Knauss \emph{et~al.}, ``{Towards State-of-the-Art and Future Trends in
  Testing of Active Safety Systems},'' in \emph{Proceedings of International
  Workshop on Software Engineering for Smart Cyber-Physical Systems
  (SEsCPS'16)}, 2016, pp. 36--42.

\end{thebibliography}

\end{document}